\documentclass[conference]{IEEEtran}

\usepackage{fancyhdr}

\fancypagestyle{icecsfirst}{
    \fancyhf{}

}

\IEEEoverridecommandlockouts

\usepackage{subcaption}
\usepackage{comment}
\usepackage[percent]{overpic}
\usepackage{cite}
\usepackage{tabularx}
\usepackage{stfloats}
\usepackage{amsmath,amssymb,amsfonts}
\usepackage[hidelinks,urlcolor=blue]{hyperref} 
\usepackage{graphicx}
\usepackage{textcomp}
\usepackage{xcolor}
\usepackage{gensymb}
\usepackage{algorithm}
\usepackage{algorithmic}
\graphicspath{{figures/}}
\def\BibTeX{{\rm B\kern-.05em{\sc i\kern-.025em b}\kern-.08em
    T\kern-.1667em\lower.7ex\hbox{E}\kern-.125emX}}

\begin{document}

\title{Real-time Tunable Probability Distributions From Perimeter-Gated SPAD TRNGs

\thanks{This material is based on work supported by the National Science Foundation under Grant No. 2442346. Any opinions, findings, and conclusions or recommendations expressed in this material are those of the authors and do not necessarily reflect the views of the National Science Foundation.}
}

\author{\IEEEauthorblockN{Yanbing Icey Chen, Md Sakibur Sajal, and Marc Dandin}
{Department of Electrical and Computer Engineering},\\
{Carnegie Mellon University,}
{Pittsburgh, Pennsylvania, 15213, USA}\\
\\
{email: mdandin@andrew.cmu.edu}   }


\maketitle
\thispagestyle{icecsfirst}

\begin{abstract}

True random number generators (TRNGs) based on the dark noise of perimeter-gated single-photon avalanche diodes (pg-SPADs) naturally produce outputs that follow a tunable Bernoulli distribution. For applications requiring non-uniform random numbers, multiple such units can be coupled, or the inherent Poisson statistics of the accumulated dark events from a single pg-SPAD can be exploited. In this work, we present the principles for extending pg-SPAD-based TRNGs from the uniform to the non-uniform regime and investigate the impact of the perimeter-gate voltage on the underlying probability distributions. The proposed concepts were experimentally validated using a prototype $\mathbf{64 \times64}$ pg-SPAD chip, fabricated in a standard $\mathbf{0.35~\mu m}$ CMOS process. Collectively, we showed that common non-uniform distributions such as Poisson, exponential, and approximated Gaussian can be achieved and tuned in real-time using pg-SPAD TRNGs. 
\end{abstract}


\section{Introduction}

Random number generation is a critical component of cryptography, data security~\cite{Seyhan2022ClassificationTaxonomy}, and emerging stochastic computing systems~\cite{tye2020}. In security-sensitive applications, random numbers (RNs) must be unpredictable, meaning unbiased and uniformly distributed. True random number generators (TRNGs) address the requirement for unpredictability by deriving entropy from nondeterministic physical processes. However, the raw outputs of physical entropy sources are often biased or non-uniform, requiring additional debiasing and post-processing circuitry to produce high-quality uniform RNs~\cite{Sajal2024TrueDiode}. 

Despite the conventional focus on uniform RNs, many applications, including Monte Carlo simulations, autonomous localization, and Bayesian machine learning, require RNs following specific non-uniform distributions~\cite{tye2020}. Conventionally, non-uniform RNs are generated by transforming uniformly distributed RNs using methods such as inverse transform sampling and acceptance--rejection sampling, which can incur significant speed and energy overhead~\cite{tye2020}. Recent hardware approaches instead exploit nonlinear device characteristics~\cite{tye2020}, combine stochastic Bernoulli bits~\cite{Zhang2024PDFConfigurableTRNGwMTJ}, or transform physically generated Gaussian RNs~\cite{osc2023} to generate target non-uniform distributions. These advances reflect a broader trend toward generating configurable probability distributions directly in hardware, with reduced computational complexity and runtime overhead.

\begin{figure}[tb]
    \centering
    \includegraphics[width=\linewidth]{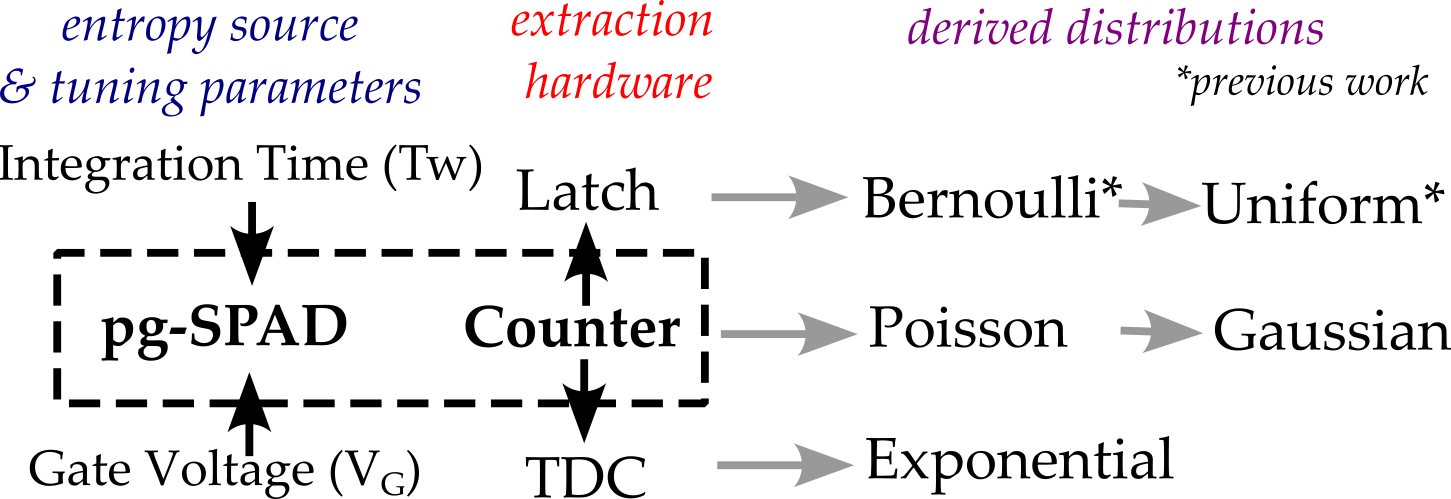}
    \vspace{-10pt}
    \caption{Tunable probability distributions extraction flow using pg-SPADs. The dashed box represents the prototype chip.}
    \vspace{-15pt}
    \label{fig:trng}
\end{figure}

Single-photon avalanche diode (SPAD) dark events have long been exploited as physical entropy sources for true random number generation~\cite{Tawfeeq2009ACounts}. Recent research in CMOS implementations continues to advance this approach toward more compact and integrated systems~\cite{calmonDCRTRNG}. Operated above breakdown in Geiger mode, SPADs amplify randomly generated carriers into detectable avalanche events, whose occurrence rate is characterized by the dark count rate (DCR). A perimeter-gated SPAD introduces an additional perimeter gate that modulates edge-dominated carrier generation and thereby provides electrical control over the DCR. This tunability was exploited in our prior work to control the output bias within a limited silicon area, enabling uniform RN generation that passed a set of NIST randomness tests~\cite{Sajal2024TrueDiode}. Here, we extend beyond uniform RN generation by extracting different statistical representations of the same underlying dark event processes to directly generate multiple RN distributions, as shown in Fig.~\ref{fig:trng}. The perimeter gate-controlled DCR further provides real-time electrical tunability of the resulting distribution parameters.

In this work, we used a pg-SPAD array with an on-chip 12-bit digital counter and implemented a time-to-digital (TDC) converter in software to generate RNs following tunable Poisson, Gaussian, and exponential distributions. As such, the following sections describe the principle, implementation, and evaluation of generating and tuning these distributions. We also discuss the limitations of the current implementation and possible solutions to address them.


\section{System Implementation}
\subsection{Theoretical Basis}
The dark events of a pg-SPAD can be modeled as a homogeneous Poisson point process with rate $\lambda$ equal to the DCR~\cite{spadmodel2023}. As demonstrated in our previous work, the DCR decreases exponentially with the perimeter-gate voltage magnitude~\cite{Sajal2024TrueDiode},
\begin{equation}
\label{eq:lambda}
\lambda(V_G)=\lambda_0 e^{-\alpha |V_G|},
\end{equation}
where $\lambda_0$ and $\alpha$ are device-dependent parameters.
Our previously reported binary TRNG represents a special case of this underlying process, where the occurrence of dark events within a fixed observation window $T_w$ is converted into an unbiased binary output through appropriate selection of $V_G$ and $T_w$ ~\cite{Sajal2024TrueDiode,Sajal2025BiasVariationCompensation}.

In this work, different probability distributions are obtained separately from the same underlying Poisson process by altering the observation mechanism and operating conditions of the pg-SPAD. Specifically, counting events within a fixed observation window produces a Poisson distribution, measuring the time interval between consecutive events yields exponentially distributed RNs, and accumulating sufficiently large event counts approximates a Gaussian distribution.

In particular, for a fixed observation window $Tw_i$, the number of detected dark events, $N_i$, follows a Poisson distribution,
\vspace{-3pt}
\begin{equation}
\label{eq:poisson}
N_i \sim \mathrm{Poisson}\!\left(\lambda(V_G)Tw_i\right),
\end{equation}
or equivalently,
\vspace{-5pt}
\begin{equation}
\Pr(N_i=k)=\frac{(\lambda(V_G)Tw_i)^k
e^{-\lambda(V_G)Tw_i}}{k!}.
\end{equation}
Likewise, the inter-event time (IET), denoted by $T_i$, follows an exponential distribution,
\begin{equation}
T_i
\sim
\mathrm{Exp}\!\left(\lambda(V_G)\right),
\end{equation}
with probability density function
\begin{equation}
f_{T_i}(t)
=
\lambda(V_G)e^{-\lambda(V_G)t},
\qquad t\ge0.
\end{equation}
Finally, when the observation window is sufficiently long such that the expected event count satisfies
$\lambda(V_G)Tw_i \gg 1$, the Poisson counting distribution can be approximated by a Gaussian distribution according to the central limit theorem,
\begin{equation}
\label{eq:gauss}
N_i
\overset{\mathrm{approx.}}{\sim}
\mathcal{N}
\!\left(
\lambda(V_G)Tw_i,
\lambda(V_G)Tw_i
\right).
\end{equation}

Equations \ref{eq:poisson} to \ref{eq:gauss} show that the parameters, \textit{i.e.}, the mean and the variance of the respective distributions can be continuously programmed through the perimeter-gate voltage, given a fixed observation window.

\begin{figure}[tbh]
    \centering
    \includegraphics[width=\linewidth]{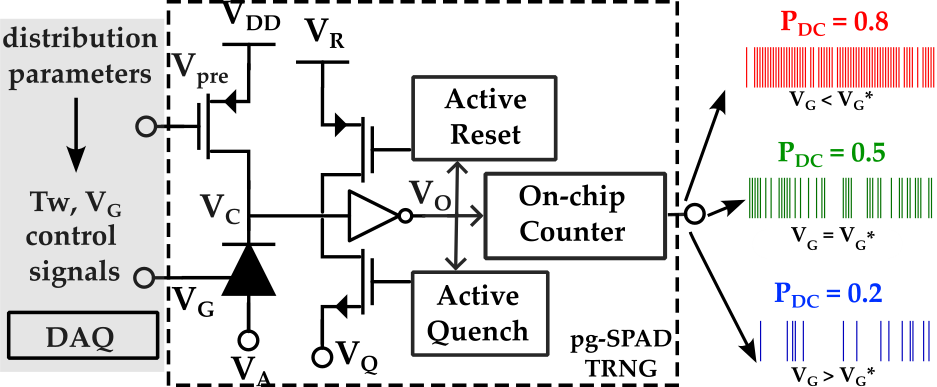}
    \vspace{-10pt}
    \caption{Simplified schematic of the dark count-based TRNG using a perimeter-gated SPAD (pg-SPAD). The dark count probability ($P_{DC}$) can be actively tuned using the gate voltage.} 
    \vspace{-5pt}
    \label{fig:sch}
\end{figure}

\subsection{Hardware/Software Implementation}

A detailed description of the pg-SPAD chip and its operation can be found in ref.~\cite{Sajal2022Perimeter-GatedProbability}. For completeness, we briefly describe the operation here in reference to Fig.~\ref{fig:sch}. The pg-SPADs continuously produce stochastic avalanche events due to dark-carrier generation. An on-chip counter accumulates events from a selected pg-SPAD over a defined observation window. An off-chip data-acquisition system then samples and records the counter output at a sampling frequency ($f_s$), from which the event-count statistics are extracted. In a free-running mode, an active quench and reset circuit respectively quench the avalanche current and reset the pg-SPAD for the following detection. A precharge device ensures the Geiger mode operation at the beginning of a data collection session. The resulting event stream is used as the common entropy source for all RVs considered in this work. Specifically, event counts within a fixed observation window $T_w$ are used to generate Poisson and Gaussian RNs, while inter-event times, computed from the differences between consecutive spike indices divided by $f_s$, yield exponentially distributed RNs. The perimeter-gate voltage $V_G$ serves as the primary control knob for real-time electrical tuning of the distribution parameters by modulating the underlying dark event rate $\lambda(V_G)$. This tuning requires no hardware modification, while different distribution families are obtained through simple software-based observation of the same event stream.

\section{Experimental Observations}

To evaluate the agreement between the measured data and the assumed probabilistic models, visual and quantitative analyses were performed for each distribution type. Empirical histograms were compared with the corresponding fitted theoretical distributions whose parameters were calculated using maximum likelihood estimation (MLE). In addition, distribution-specific statistical metrics \textit{e.g.}, variance-to-mean ratio (VMR) for assessing dispersion in Poisson counts, skewness and kurtosis for evaluating Gaussian behavior were computed to quantify model consistency.

\vspace{-2pt}
\subsection{Poisson Distribution}
\begin{figure}[h]
\vspace{-5pt}
    \centering
    \includegraphics[width=0.95\linewidth]{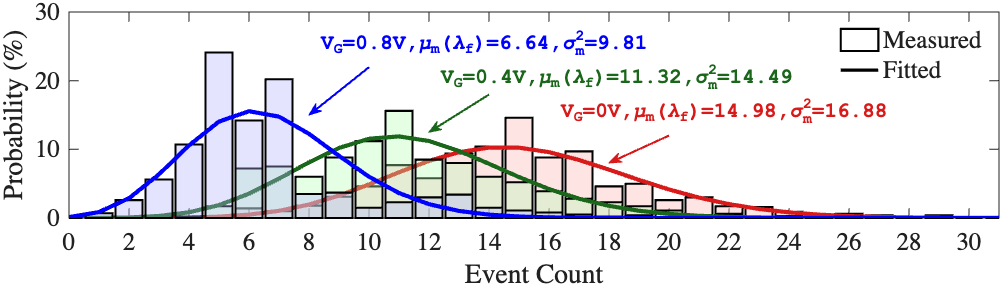}
    \vspace{-3pt}
    \caption{Poisson distribution at different $V_G$, with measured histograms and fitted curves. Here, $T_{w} = 2~ms$.} 
    \label{fig:poisson_hist}
    \vspace{-10pt}
\end{figure}
Figure~\ref{fig:poisson_hist} compares the measured histograms of windowed event counts with the corresponding ideal Poisson probability mass functions (PMFs), where the fitted parameter $\lambda_f$ equals the sample mean $\mu_m$. As $V_G$ increases from 0 to 0.8 V in 0.4 V steps, the mean count per window decreases, consistent with the dark-operation principle summarized in Fig.~\ref{fig:sch}. The measured VMR ranges from 1.13 to 1.48, remaining close to the ideal value of 1, with the slight overdispersion attributable to afterpulsing~\cite{spadmodel2023}. This small residual deviation from the ideal Poisson assumption can be reduced through dead-time and afterpulsing correction, or modeled explicitly when non-ideal counting behavior is itself of interest~\cite{Bektashi_2022_poisson_dispersion}.

\subsection{Exponential Distribution}

\begin{figure}[h]
\vspace{-5pt}
    \centering
    \includegraphics[width=0.95\linewidth]{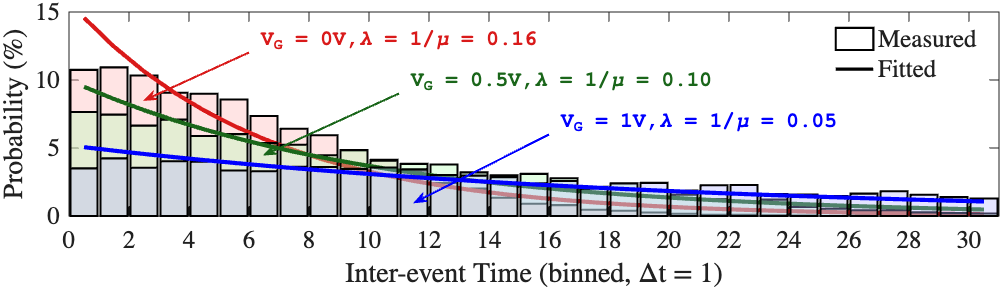}
    \vspace{-3pt}
    \caption{Exponential distribution at different $V_G$, with measured histograms and fitted curves. Here, $T_{w} = 2~ms$.} 
    \label{fig:exponential_hist}
    \vspace{-5pt}
\end{figure}
The corresponding ideal exponential PDFs were constructed using the fitted rate parameter $\lambda = 1/\mu$. As $V_G$ increases, the mean inter-event time increases as the event rate $\lambda$ decreases accordingly as seen in Fig.~\ref{fig:exponential_hist}, which shows close agreement over the main probability range. Although the probability of very small inter-event times falls below the ideal exponential prediction, this behavior is attributed to the finite dead time and recovery process of the pg-SPAD and readout circuitry, which reduces the likelihood of observing closely spaced consecutive events~\cite{spadmodel2023}.

\subsection{Gaussian Distribution}

\begin{figure}[h]
\vspace{-5pt}
    \centering
    \includegraphics[width=0.95\linewidth]{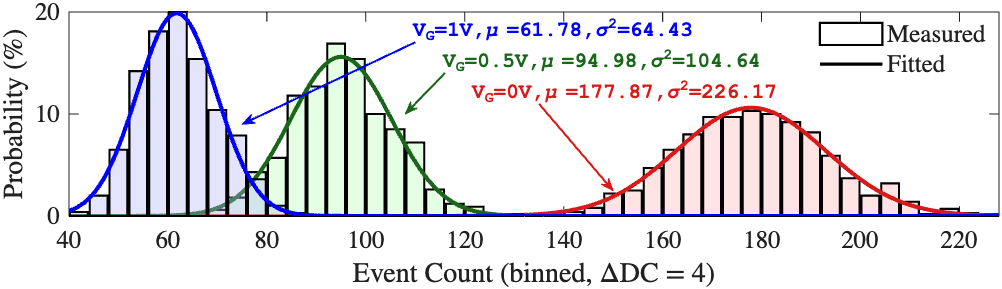}
    \vspace{-3pt}
    \caption{Gaussian distribution at different $V_G$, with measured histograms and fitted curves. Here, $T_{w} = 5~ms$.} 
    \label{fig:gaussian_hist}
    \vspace{-5pt}
\end{figure}
\vspace{-1pt}
For a longer observation window, the measured event-count histograms become closer to Gaussian PDFs constructed using the sample mean $\mu$ and variance $\sigma^2$, consistent with the central limit theorem. For visual clarity, the event counts were grouped into bins before plotting (see Fig.~\ref{fig:gaussian_hist}). As $V_G$ increases from 0 to 1 V in 0.5 V steps, both $\mu$ and $\sigma^2$ decrease accordingly, continuing the trend observed in Fig.~\ref{fig:sch}. The measured skewness values are 0.183, 0.198, and 0.335, and the corresponding kurtosis values are 2.989, 2.947, and 2.967, respectively, remaining close to the ideal Gaussian values of 0 and 3, respectively~\cite{Hatem2022_ks_sk}.

\section{Discussion}
The experimental results demonstrate that different observation operators applied to the same underlying dark-event process can produce RNs consistent with the intended probability distributions. The randomness of the underlying binary entropy source at the unbiased operating point ($P_{DC}=0.5$) has been previously validated using the NIST test suite~\cite{Sajal2024TrueDiode,Sajal2025BiasVariationCompensation}. The present results extend this characterization from binary randomness to the fidelity of the derived probability distributions, with their statistical parameters electrically tuned through the perimeter-gate voltage.

\begin{figure*}
    \centering
    \includegraphics[width=0.95\linewidth]{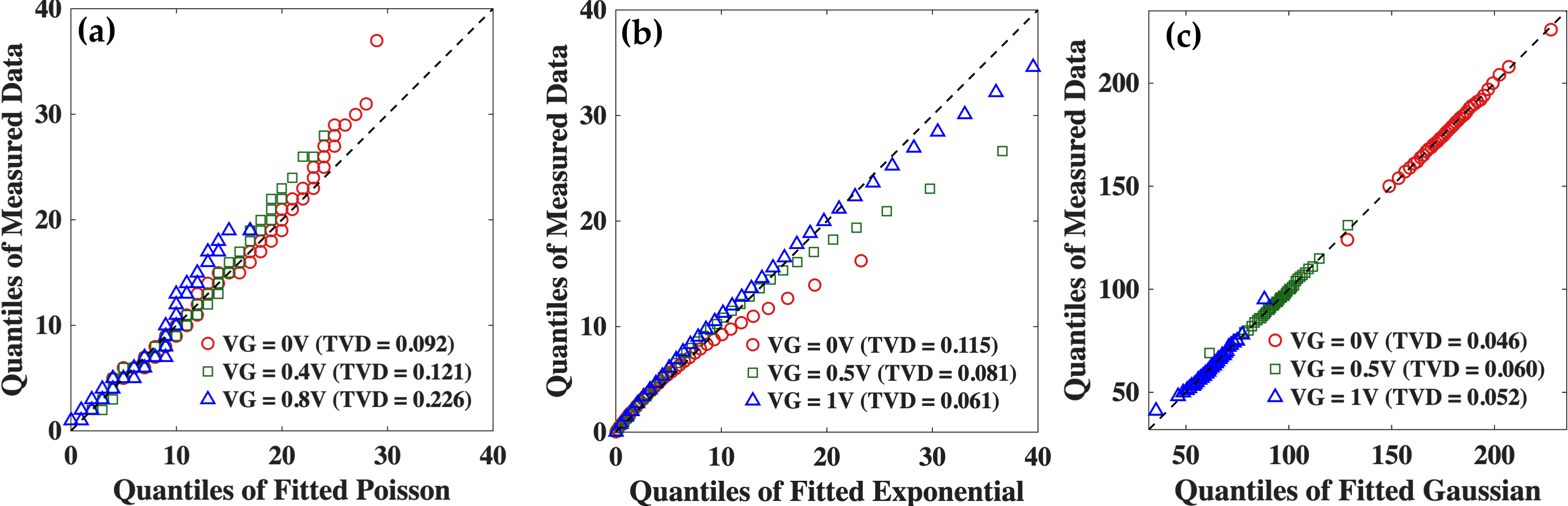}
    \caption{ Q--Q plots comparing the measured and theoretical quantiles for the (a) Poisson, (b) exponential, and (c) Gaussian distributions. Closer agreement with the reference line $y=x$ indicates a better match between the empirical and theoretical distributions. The corresponding TVD values are reported in the legends.}
    \vspace{-10pt}
    \label{fig:qq}
\end{figure*}

The deviations observed in the Q--Q analysis also provide insight into device-level nonidealities. The Poisson distribution in Fig.~\ref{fig:qq}(a) exhibits slight upward deviation at higher quantiles, consistent with the mild overdispersion observed from the VMR. The exponential distribution in Fig.~\ref{fig:qq}(b) shows greater deviation from the ideal model, which may arise from finite dead time, recovery dynamics, and other temporal nonidealities in the pg-SPAD and readout chain~\cite{Ziarkash2018Afterpulsing}. In comparison, the Gaussian distribution in Fig.~\ref{fig:qq}(c) maintains close agreement with the theoretical quantiles, suggesting that aggregation over larger event counts reduces the influence of individual event-level nonidealities.

\section{Limitations and Future Work}
We have shown through previous sections that the proposed framework enables direct generation of multiple distributions with programmable parameters. While these results demonstrate the feasibility and versatility of the approach, several limitations remain and motivate future investigation.
\subsection{Environmental Robustness:}
Temperature variation and device aging can alter the pg-SPAD dark-count rate $\lambda$ and consequently shift the parameters of the generated distributions~\cite{sajal2026modelingdarkcountprobability}. However, the perimeter-gate voltage provides a direct calibration knob to tune $\lambda$ and compensate for such drifts, enabling partial stabilization of the distribution parameters. While this work does not experimentally characterize long-term stability or closed-loop calibration, the availability of gate-voltage tuning suggests a practical pathway for environmental compensation. Future work will investigate these effects and develop appropriate calibration strategies.

\subsection{Statistical Dependence:}
The distributions demonstrated in this work were generated from different pg-SPAD pixels and separate acquisition runs to establish the feasibility of multiple distributions. Consequently, cross-distribution correlation was not directly evaluated in the present experiments. For future implementations requiring simultaneous generation of multiple distributions, correlations between random variable streams may arise from overlapping event records, inter-pixel coupling, or shared readout circuitry. Future work will quantify temporal and cross-distribution correlations and investigate nonoverlapping acquisition windows or separate pg-SPAD pixels for generating independent random-variable streams.
\vspace{-5pt}
\subsection{System-Level Performance:}
A complete characterization of power, area, and throughput is beyond the scope of the present device-level demonstration. The achievable generation rate is currently constrained by the intrinsic dark event rates, available on-chip counters, and off-chip acquisition bandwidth. Future implementations will explore photon-induced events for higher generation rates, parallel on-chip architectures, and integrated distribution extraction to enable systematic evaluation of energy per sample, area, and distribution-specific throughput.

\section{Conclusions}
This work demonstrates that a pg-SPAD can serve as a reconfigurable entropy engine capable of generating random numbers from multiple probability distributions rather than merely a uniform distribution. Switching among distributions is achieved by applying different observation mechanisms to the same underlying dark-event process, while real-time tuning of their statistical parameters is enabled by varying the perimeter-gate voltage. Unlike conventional approaches that synthesize target distributions by computationally transforming uniform random numbers, the proposed approach directly exploits the native statistics of the physical entropy source, reducing the need for complex post-processing algorithms and dedicated hardware. Despite nonidealities associated with SPAD afterpulsing and dead-time effects, the measured distributions show close agreement with the corresponding theoretical models. These results establish the pg-SPAD as a promising platform for probability-native sensing and computing applications.


\bibliographystyle{IEEEtranDOI}

\bibliography{main}

\end{document}